\documentclass[a4paper,12pt]{article}
\usepackage[utf8]{inputenc} 
\usepackage[T1]{fontenc}
\usepackage{graphicx,caption}
\usepackage{amsmath}

\usepackage{amsfonts} 
\usepackage{amsthm} 
\usepackage{amssymb} 
\usepackage{array}
\usepackage{braket}
\usepackage[left=3cm,right=3cm,top=3cm,bottom=3cm]{geometry}
\usepackage{url}
\usepackage[dvipsnames]{xcolor}
\usepackage{tikz}
\usetikzlibrary{shapes,arrows,positioning}
\usepackage[numbers]{natbib}

\usepackage{lipsum}
\usepackage{calligra}
\usepackage[mathscr]{euscript}
\DeclareMathAlphabet{\mathcalligra}{T1}{calligra}{m}{n}

\definecolor{vert}{rgb}{0.3,0.7,0.3}
\definecolor{bleu}{rgb}{0.4,0.6,1}
\definecolor{violet}{rgb}{0.2,0,0.7}

\newcommand{\ee}{{\rm e}}

\newcommand{\ii}{{\rm i}}
\newcommand{\dd}{{\rm d}}

\title{From the geometry of Foucault pendulum \\ to the topology of planetary waves}

\author{ Pierre Delplace and Antoine Venaille\\ 
\footnotesize
  \it Univ Lyon, Ens de Lyon, Univ Claude Bernard, CNRS,  \\ \footnotesize \it Laboratoire de Physique, F-69342 Lyon, France}

\date{}

\begin{document}

\maketitle

\section*{Abstract}
{\bf
The physics of topological insulators makes it possible to understand and predict the existence of unidirectional waves trapped along an edge or an interface. In this review, we describe how these ideas can be adapted to geophysical and astrophysical waves. We deal in particular with the case of planetary equatorial waves, which highlights the key interplay between rotation and sphericity of the planet, to explain the emergence of waves which propagate their energy only towards the East. These minimal ingredients are precisely those put forward in the geometric interpretation of the Foucault pendulum. We discuss this classic example of mechanics to introduce the concepts of holonomy and vector bundle which we then use to calculate the topological properties of equatorial shallow water waves.
}

\section*{Résumé}
{\bf
La physique des isolants topologiques permet de comprendre et prédire l'existence d'ondes unidirectionnelles piégées  le long d'un bord ou d'une interface. Nous décrivons dans cette revue comment ces idées peuvent être adaptées aux ondes géophysiques et astrophysiques. Nous traitons en particulier le cas des ondes équatoriales planétaires, qui met en lumière les rôles clés combinés de la rotation et de la sphéricité de la planète pour expliquer l'émergence d'ondes qui ne propagent leur énergie que vers l'est. Ces ingrédients minimaux sont précisément ceux mis en avant dans l'interprétation géométrique du pendule de Foucault. Nous discutons cet exemple classique de mécanique pour introduire les concepts d'holonomie et de fibré vectoriel que nous utilisons ensuite pour le calcul des propriétés topologiques des ondes équatoriales en eau peu profonde.
}

\section{The renewal of topological waves}

As recalled by Michael Berry \cite{Berry_miraculous}, the investigation of the topological properties of waves started  during the ''miraculous 1830s'' with the discovery of their singularities~: the singularity of the intensity explaining the emergence rainbows; the singularity of the phase, as amphidromic points\footnote{Amphidromic points  are phase singularity where tidal amplitude must vanish.} discovered at that time in the North sea; and the singularity of the polarization whose theoretical prediction led to the observation of the conical refraction in optics.
Topology of waves was enriched in the late seventies, for instance with the discovery of wavefront dislocations of water waves that emerge when scattered by a vortex, thus providing a classical analog to the quantum Aharonov-Bohm effect \cite{Berry80}. This example anticipates how topological properties of quantum wave functions may inspire the search for novel topological properties of classical waves. Indeed, for the last ten years, it was realized that topological properties similar to that of the integer quantum Hall effect and of the recently discovered topological insulators could be engineered in metamaterials with classical waves of various kinds, from optics \cite{TopoPhot} to mechanics \cite{nash2015topological,Susstrunk2015} and acoustics \cite{Zhang:2018aa,Souslov:2017aa}. %
These topological properties are related to phase singularities of the complex eigenstates of the system in a parameter (or reciprocal) space, and translate in real space as the existence of trapped boundary modes that can be used to guide energy, through the celebrated \textit{bulk-boundary correspondence} \cite{HatsugaiPRL1993}. These confined states are often referred to as  \textit{topological modes}.

Coincidentally, it was during this same "miraculous" 1830 decade that Gaspard-Gustave Coriolis formalised the celebrated inertial force that nowadays bears his name. A spectacular manifestation of this force is revealed by the slow deviation of the Foucault pendulum, an effect that can precisely be apprehended with a geometrical approach \cite{Berry5years}. As we shall see below, the Coriolis force is also involved in a singularity of geophysical fluid waves, associated to the twisting of eigenmodes around degeneracy points in their dispersion relation. This topological property is closely related to the geometric interpretation of the Foucault pendulum. It manifests itself through the existence of peculiar equatorially trapped eastward oceanic and atmospheric waves, which bear strong formal similarities with  boundary states of a topological material. It was  shown over the last few years that topological waves are indeed ubiquitous in natural systems, with application to equatorial dynamics \cite{delplace2017topological}, astroseismology \cite{perrot2019topological}, plasma \cite{parker2020topological}, or active matter  \cite{shankar2017topological,souslov2019topological}.

We review here the recent input of topological tools inherited from topological insulators to these geophysical and astrophysical waves. We put emphasis on equatorial waves, which highlight the crucial role of Earth rotation and curvature.  We propose to use the Foucault pendulum  as a starting point to introduce key notions of geometrical properties induced  by a rotating planet, and then use these tools to address the topology of equatorial waves.

\section{Coriolis force-induced geometrical effects}

\subsection{Coriolis force on Earth}

Coriolis force is an inertial force perceived by an object of velocity $\mathbf{v}$ in a rotating frame of reference.  Its effect  is to deviate the object's trajectory in a direction perpendicular to both $\mathbf{v}$ and the rotation vector $\boldsymbol{\Omega}$, i.e. it reads  $\mathbf{F}_c = 2 M \mathbf{v} \times \boldsymbol{\Omega} $
where $M$ is the mass of the object. 
Owing to its rotation, Earth naturally induces a Coriolis force for an observer at the surface of the planet. However, its effect is clearly negligible at the scale of a human being: to experience the Coriolis force of only 1\% amplitude of our weight, we should run at about $2\,500\, km/h$...\footnote{$2 v \Omega/g = 10^{-2}$ with $g=10$  m.s$^{-2}$ and $\Omega =2\pi/(24\times 60 \times 60)$ s$^{-1}$}. This is due to the small value of the angular velocity of our planet. One can experience this force in  inertial carousels where the visitors are invited to pass each others a ball; they fail because of the strong deviation induced by the fast rotation of the carousel. 

The Coriolis force cannot be neglected anymore 
 when the motion occurs over a time scale $T$ comparable to the period of rotation, i.e. when $1/(\Omega T)$ is of order one or smaller. 
This is well illustrated by the celebrated Foucault pendulum, whose vertical oscillation plane significantly deviates from its initial orientation when oscillating long enough.
Similarly, slow motion of typical velocity $U$ in ocean currents and atmospheric winds are strongly affected by the Coriolis force  at the scale $L$ of a planet, that is again when  $U/(\Omega L)$ is of order one or smaller.
In the case $U/(\Omega L)\ll 1$, a striking manifestation of Earth rotation is that currents or winds blow along pressure lines, as horizontal  momentum equations are dominated by a balance between Coriolis and pressure forces. This property is used  to draw midlatitude weather maps, where cyclones and anticyclones are visualized by using isobars. The long time behavior of the Foucault pendulum and the slow motion of winds or oceanic currents at the scale of a planet thus both reveal the influence of Earth rotation. As we explain below, geometrical tools are also in both cases particularly useful to understand central aspects of their dynamics. It is necessary for that purpose to introduce a second  key ingredient: Earth's curvature.

Because of Earth's roundness, the effect of the Coriolis force depends on the latitude. Dynamics at the poles is very much like in the carousel picture, since the plane tangent to the sphere is in that case perpendicular to the angular rotation vector $\boldsymbol{\Omega}$. The situation is quite different at the Equator, as the local tangent planes contain the planet rotation vector $\boldsymbol{\Omega}$.
Standard descriptions of the Foucault pendulum and planetary waves rely on two assumptions: the motion it nearly horizontal, and in-plane component of the angular rotation $\boldsymbol{\Omega}$ are neglected. In the Foucault pendulum case, both assumptions are justified in a small amplitude limit. In the geophysical case, this can be justified by considering the limit of vanishing aspect ratio between the vertical fluid layer thickness and horizontal scale of motion \cite{VallisBook}. In both cases, only the horizontal components of the Coriolis force matters,  and  the effect of planet rotation is encoded into the \textit{Coriolis parameter}
\begin{equation}
f=2\, \Omega \sin \theta,   \label{eq:def_coriolis_parameter}
\end{equation}
with $\theta$ the latitude. This parameter is twice the projection of the planet rotation vector on the local vertical axis ($2\boldsymbol{\Omega} \cdot \hat{\mathbf{n}}$). 
The Coriolis parameter increases from the South pole  ($\theta=- \pi/2$) to the North pole  ($\theta=\pi/2$) , and vanishes at the Equator ($\theta=0$) The latitudinal variations of the Coriolis parameter $f$ is the second key ingredient to understand important aspects of Foucault pendulum dynamics and geophysical flows at planetary scale. In fact, it  plays a central role in the geometrical arguments exposed in the following sections.

\subsection{A geometrical look at the Foucault pendulum}

In January 1851, L\'eon Foucault observes in his basement  the slow but ''irresistible'' clockwise deviation of the oscillation plane of a two meter long pendulum hung from the ceiling, thus subtly revealing Earth's rotation \cite{Foucault}. Its demonstration was reproduced a month later at the Observatoire de Paris and the next month at the Pantheon in Paris, this time with a steel wire of $67$m and a globe of $28$kg, in order to make visible the Earth's rotation to everyone \cite{sommeria2017foucault}. 

This slow deviation of the oscillatory plane can be directly inferred from a standard procedure, when taking into account the Coriolis force acting on the pendulum. 
  In the small amplitude limit, one can safely neglect the vertical velocity of the pendulum, and the horizontal trajectory can then be written in the compact form $z(t)=x(t)+\ii y(t)$. Integration of Newton equations then leads to $z(t)=\alpha \ee^{-\ii (\omega - f/2)t} + \beta \ee^{-\ii (-\omega + f/2)t}$, where $\omega$ is the pulsation of the pendulum and $f$ is the Coriolis parameter defined in Eq. (\ref{eq:def_coriolis_parameter}),  and where the constants $\alpha$ and $\beta$ depend on the initial conditions. In the limit $ f \ll \omega$,  the vertical oscillation plane of the pendulum slowly deviates clockwise in the Northern hemisphere ($f>0$).

%

\begin{figure}[h]
\center
\includegraphics[width=13.5cm]{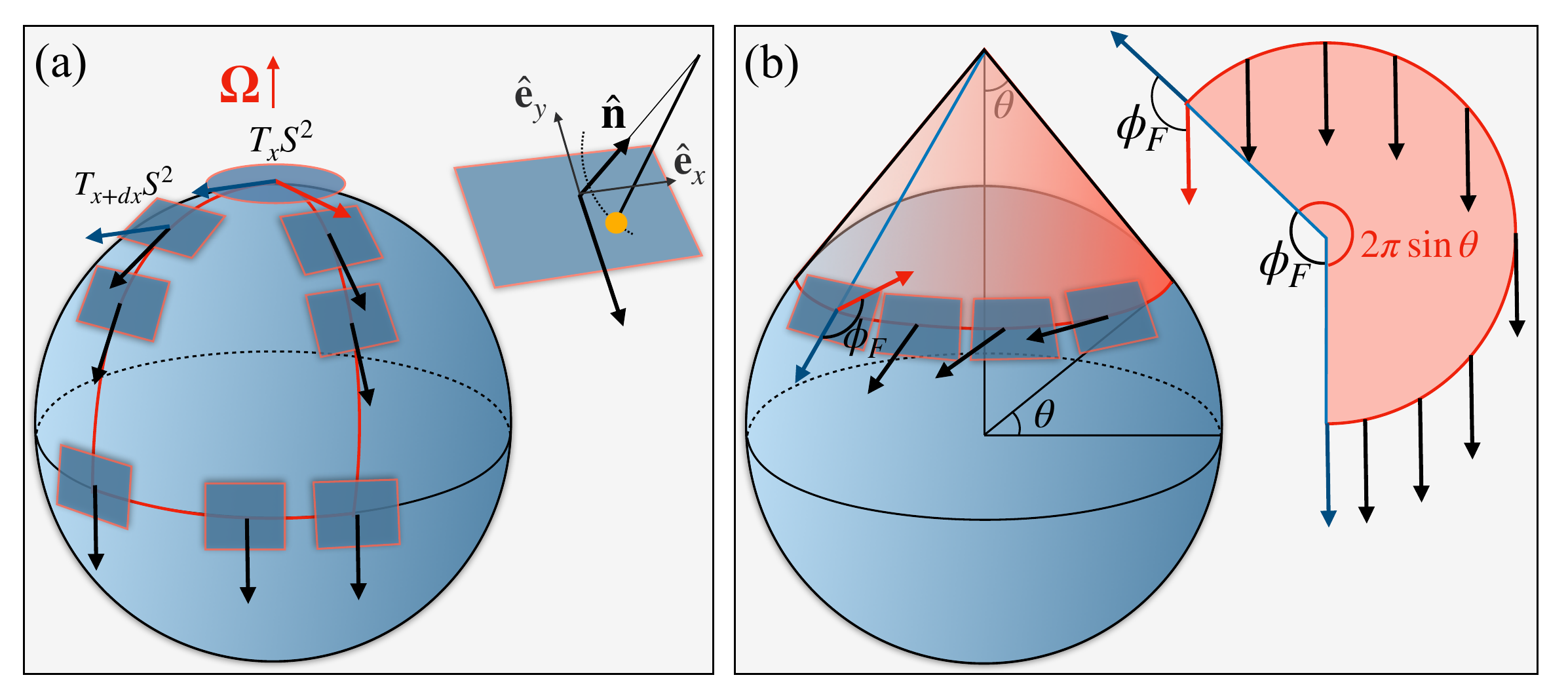}
\caption{(a) Parallel transport of a tangent vector along geodesics. After a close path, the transported vector (red) differs from the initial one (blue) by an angle. Inset: Foucault pendulum; its trajectory  (dashed), defines the black arrow once projected onto the tangent plane. (b) Parallel transport of a tangent vector along a close path (red) at fixed latitude $\theta$, that continuously rotates in its tangent plane. Inset: Unfold and flattened red cone.}
 \label{fig:parallel}
\end{figure}

The deviation of the Foucault pendulum can also be apprehended from elegant geometrical considerations in an inertial frame of reference \cite{Berry5years}. This approach does not require any knowledge of the Coriolis force, and is motivated by the slow motion of Earth rotation relative to  that of the pendulum oscillations. This scale separation is referred to as the \textit{adiabatic}\footnote{This adiabatic limit is not related to heat transfers in the system. } limit. Locally, the pendulum trajectory  is described  in a tangent plane to Earth, as sketched in figure \ref{fig:parallel}(a).
The pendulum inertia tends to simply translate the pendulum trajectory parallel to itself in space. Of course, the pendulum is retained on Earth because of its weight and of the tension of the wire it is attached to. Those constraints guarantee that the pendulum trajectory always remains in a  plane tangent to Earth. After one period of pendulum oscillation, Earth has \textit{adiabatically} rotated: the pendulum  trajectory has barely changed, but it does not belong any longer to the same tangent plane to the sphere in the inertial frame of reference.

 We are then left with the question of how its projected trajectory -- that we shall represent with a tangent vector to Earth's surface --  can remain \textit{at best} parallel to itself when continuously moved from a tangent plane to another, that are arranged along a curve on the surface of Earth, more specifically at fixed latitude? This question is known in Mathematics as the one of \textit{parallel transport} of vectors, and was in particular worked through by Tullio Levi-Civita at the beginning of the XXth century in parallel to Einstein's general relativity. 
  
 In the pendulum case, parallel transport gives a procedure to displace a vector in $\mathbb{R}^3$ from a tangent plane $T_xS^2$ to the sphere $S^2$ at a point $x$ to another tangent plane $T_{x+\dd x}S^2$. This procedure tells that, when parallel transporting a vector along a curve, the rate of change of this vector must be normal to the surface, as it is so in the Euclidian planar geometry.
Vector fields of parallel transported tangent vectors are illustrated in figure \ref{fig:parallel} along different curves. These are intuitively obtained when the curve corresponds to a geodesics, e.g. along the Equator or along a Meridian  (see figure \ref{fig:parallel}(a)). Perhaps less intuitively, a vector parallel-transported along an arbitrary cycle generally does not return identical to itself, but makes an angle with the initial vector. 

As for the Foucault pendulum, we only care about parallel transport of vectors along a longitude. The particular case of the Equator ($\theta=0$), that is a geodesic, follows from the discussion above~: in that case, a local observer contemplating the pendulum in the tangent planes does not see any deviation of its oscillations. It is thus worth stressing that the deviation of the Foucault pendulum is a consequence of the interplay between Earth rotation and Earth local curvature~: for instance, there would not be any deflection on a rotating cylinder.
The deflection angle $\phi_F$ that the pendulum makes between its initial and final orientations after one day depends on the latitude $\theta$. It can be easily obtained geometrically, by noting that the collection of planes tangent to the sphere along a longitude are also tangent to a cone (see figure \ref{fig:parallel}(b)). By unfolding and flattening this cone, the parallel transport of vectors reverts to the usual and intuitive one of Euclidian geometry. Crucially, this operation requires to cut the cone, and we choose the cut position to coincide with the initial and final position of the pendulum (blue line in figure  \ref{fig:parallel} (b)). 
It becomes clear that, to be parallel transported along a longitude, a tangent vector in the Northern Hemisphere has to rotate clockwise, as observed by Foucault. Moreover, it follows from an elementary geometrical analysis that it acquires an angle $2\pi \sin \theta$ after a full cycle. After one day of oscillations, the pendulum thus makes a deflection angle $\phi_F=2\pi(1-\sin \theta)$ with its initial orientation. 

To summarize, the mismatch angle $\phi_F$ of the Foucault pendulum follows from two ingredients ~: (1) parallel transport along a longitude, that is due to the adiabatic rotation of Earth, and (2) the curvature of the surface. More generally, such a phase mismatch of a parallel transported vector over a  loop on a surface  reads
  \begin{align}
  \phi_F= \int_{\Sigma} \kappa\,  \dd S 
    \label{eq:phiF}
  \end{align}
where $\kappa$ is the gaussian curvature of the surface and $\Sigma$ is the surface enclosed by the loop.
Considering Earth as a perfect sphere of radius $R$, this curvature is simply $\kappa=1/R^2$, and one recovers the result aforementioned which is nothing but the solid angle delimited by the close path along a circle of latitude.

It has been found that the polarization of seismic shear waves travelling over the Earth slowly rotates similarly to the Foucault pendulum \cite{snieder2016seismic}. In the following we present another consequence of Earth rotation and sphericity on planetary waves that is related to a topological number.

\subsection{From  geometrical phases to topological numbers for waves}
\label{sec:geomtopo}
The phase accumulated by a system over a cycle is a generic property of vector bundles, called holonomy. A vector bundle can be seen as a continuous collection of vector spaces parametrized over a close manifold.  In Foucault pendulum case, the  collection of vector space  are tangent planes $\mathbb{R}^2$, and the base space is the sphere $S^2$. Vector bundles are mathematical objects that appear in physics for instance when a gauge freedom is involved. In quantum mechanics, the local choice of the phase of the wavefunction is an important example of gauge freedom that gives rise to $U(1)$-vector bundles. There, quantum eigenstates can acquire a geometrical phase, known as the Berry phase \cite{Berry1984, Simon}, when a periodic modulation of the system is performed adiabatically. 
There is thus a conceptual common root between the quantum holonomy, and that of the Foucault pendulum. However, the nature of the fiber bundle involved is different, and the curvatures used to describe local geometrical properties of these bundles are also different. In the Foucault pendulum case, the Gaussian curvature was used in Eq. (\ref{eq:phiF}).  In the quantum case, a different (two-form) curvature $\mathcal{F}^{(n)}(\boldsymbol{\lambda})$ is employed, called Berry curvature, which is a property of parametrized eigenstates $\psi_{n}(\boldsymbol{\lambda})$ with  $\boldsymbol{\lambda}=(\lambda_1, \lambda_2, \dots)$ a point in parameter space.
The Berry curvature is a physical observable that was measured in of quantum \cite{Jotzu:2014aa} and classical \cite{Wimmer2017} systems.  In particular, it appears in the semiclassical equations of motion of electronic wavepackets by yielding a correction to the group velocity to that predicted from the dispersion relation alone. A similar effect is currently investigated in geophysical ray tracing \cite{perez2020manifestation}.

Similarly to the phase mismatch \eqref{eq:phiF}, the Berry phase results from the integration of Berry curvature, which formally reads
 \begin{align}
  \phi_B^{(n)}= \int_{\Sigma} \mathcal{F}^{(n)} \qquad \mathcal{F}^{(n)}=\ii \left( \frac{\partial \psi_{n}^\dagger}{\partial \lambda_i} \frac{\partial \psi_{n}}{\partial \lambda_j} - \frac{\partial \psi_{n}^\dagger}{\partial \lambda_j} \frac{\partial \psi_{n}}{\partial \lambda_i}  \right) \dd \lambda_i \wedge \dd \lambda_j \ .
  \label{eq:berry}
  \end{align} 

A pedagogical model introduced by Berry to illustrate this geometrical phase consists of a quantum spin $\mathbf{\hat{S}}$ coupled to a slowly varying classical magnetic field $\mathbf{B}$  \cite{Berry1984}.  The dynamics is encoded in the Hamiltonian $\hat{H}=\mu \mathbf{B}.\mathbf{\hat{S}}$ where $\mu$ is a constant involving the gyromagnetic ratio.  Spin eigenstates are denoted by $\ket{m}$, with $m=\{-S,-S+1 ,\cdots S\}$. When the orientation of $\mathbf{B}$ is varied \textit{adiabatically}  along a close path, a spin eigenstate $\ket{m}$ acquires a Berry phase   $\phi_B^{(m)}=-m \Omega_s$  where $\Omega_s$ is the solid angle drawn by $\mathbf{B}$, by analogy with the deflection angle of the Foucault pendulum in figure \ref{fig:parallel} (b).

When $\Sigma$  consists in the entire close manifold base space, the holonomy  along its boundary is meaningless. Still, the integration of the Berry curvature is meaningful, and is actually an integer-valued topological index that counts the number of singularities of the $U(1)$-vector bundle. For instance, in Eq. \eqref{eq:phiF}, the integration of the Gaussian curvature  over an orientable close surface $\Sigma$ is the Euler-Poincaré integer number $\frac{1}{2\pi}\int_\Sigma \kappa \dd S =  2(1-g)$ that only depends on the genus $g$ of the surface. For the sphere ($g=0$), the value $2$ then obtained is the number of vortices where any tangent vector field necessarily vanishes. The integration of the Berry curvature over the base space has a quite similar meaning; it corresponds to the number of phase singularities  of the complex eigenstate. This topological number, called the (first) Chern number, was introduced in physics in 1982 to explain the unexpected remarkable quantization of the transverse conductivity of the quantum Hall effect \cite{TKNN} and was later found to predict the number of unidirectional modes propagating without dissipation along the edge of the sample \cite{HatsugaiPRL1993}.

\begin{figure}[h]
\center
\includegraphics[width=15cm]{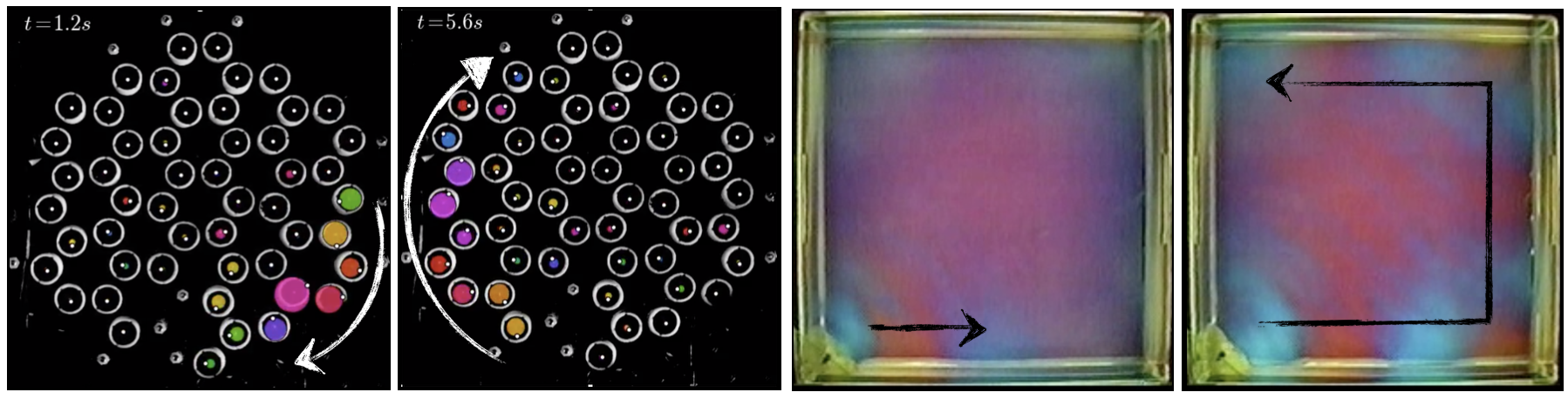}
\caption{Left: clockwise mechanical edge mode propagating in an array of coupled gyroscopes, reproduced from \cite{nash2015topological}. Colors encode the phase of the gyroscopes.  Right: laboratory realization of a coastal Kelvin wave in a rotating tank, reproduced from \cite{GFDdennou}. Colors encodes variations in the fluid layer thickness. Snapshots are taken at two successive times to visualize wave propatation. }
 \label{fig:meta}
\end{figure}

 Although the topological quantization of the conductivity is a specific property of quantum electronic wavefunctions, the existence of topologically protected unidirectional edge modes was later realized to be an ubiquitous property of wave dynamics across all fields of physics, provided that time-reversal symmetry is broken. This requirement is satisfied by the presence of a perpendicular magnetic field in the Hall effect. It must be engineered differently when dealing with classical neutral waves. The formal similarity between  Lorentz and Coriolis forces fruitfully suggests that rotating the system is a natural alternative. A nice illustration of the effect of rotation on mechanical waves was realized with a lattice of  gyroscopes  coupled by springs \cite{nash2015topological}. A regime of couplings can be found such that when a gyroscope is excited at the edge of the lattice, it generates a wave that propagates clockwise along the boundary, without spreading into the bulk, despite the abruptness of the irregular edge (figure \ref{fig:meta}). Another possibility is to rotate the entire system itself rather than its constituants. In the case of a thin fluid layer in a rotating tank, this leads to the emergence of unidirectional trapped modes that  propagate along the tank's wall (figure \ref{fig:meta}). Such waves are nothing but a lab implementation of coastal Kelvin waves, that propagate along natural coats of lakes and continents due to the Coriolis force induced by Earth's rotation \cite{VallisBook}. Note that contrary to the case of gyroscopes, and to other macroscopic experiments designed to exhibit mechanical or fluid topological waves \cite{Susstrunk2015,Souslov:2017aa}, there is no underlying lattice in the  rotating fluid case, which is not without causing important conceptual issues concerning the topological nature of these boundary waves \cite{tauber2019bulk,tauber2020anomalous,tauber_violation20}.

\subsection{Topological equatorial waves}

Equatorial atmospheric and oceanic waves are another emblematic example of unidirectional trapped modes in geophysics. The Equator around which these waves are trapped  plays a role analogous to that of a coast for the Kelvin waves, except that the Equator is not a boundary of the fluid.  Nevertheless, it can somehow be interpreted as a kind of interface between two Hemispheres where the Coriolis parameter  $f$ defined in Eq.  (\ref{eq:def_coriolis_parameter})  has opposite sign. Because time-reversal symmetry is broken by the Coriolis parameter $f$, one may expect peculiar topological waves at the interface. Indeed, he problem of a continuous system with such a smooth interface turns out to be a convenient framework to show the topological origin of confined unidirectional waves, in the sense that their number is precisely given by the Chern invariant.

\begin{figure}[h]
\center
\includegraphics[width=15cm]{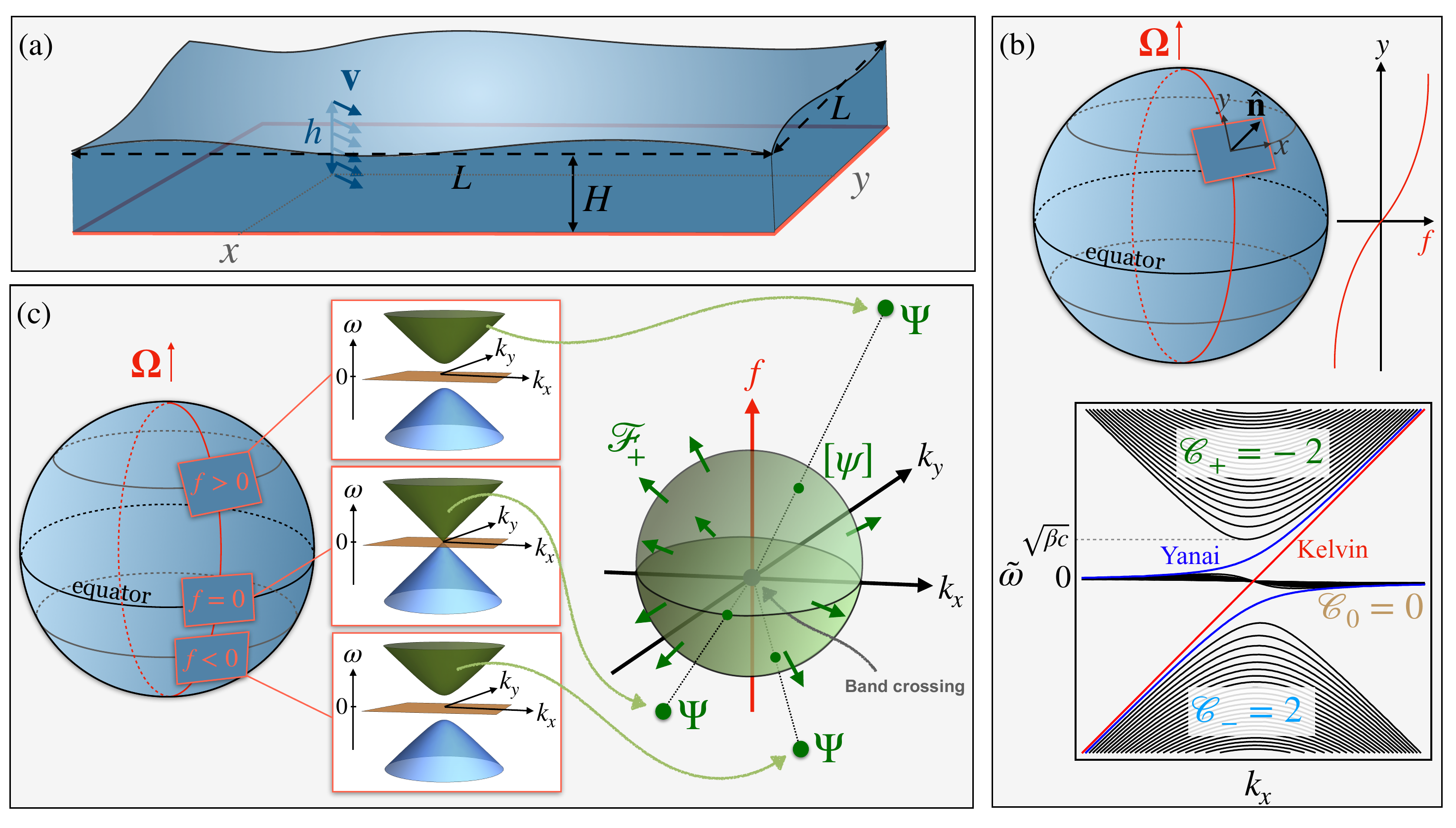}
\caption{(a) Shallow water model. Thickness $H$ is much smaller than horizontal length scale $L$. (b) Upper panel: variation of the Coriolis parameter with latitude on the sphere. Lower panel: dispersion relation of shallow water waves computed by T. Matsuno in 1966 under the beta plane approximation $f=\beta y$. Notice the presence of two unidirectional modes filling the gap between low frequency Rossby waves and high frequency Poincar\'e waves. Those are the equatorial Kelvin and Yanai waves. (c) Dispersion relation of shallow water waves with constant $f$, for different tangent planes to the rotating sphere.
The gap vanishes when $f=0$. In that case, a degeneracy point occurs at the origin in $(k_x,k_y)$ plane. The right panel is a schematic representation of the eigenmode bundles in parameter space $(k_x,k_y,f)$. The base space in a sphere surrounding the degeneracy point. The topological charge of this degeneracy point explains the emergence of equatorial Kelvin and Yanai waves.}
\label{fig:SW}
\end{figure}

Equatorial waves at planetary scale are well described by the linearized shallow water model (see figure \ref{fig:SW} (a)). This textbook model  describes an incompressible fluid whose thickness  is much smaller than the horizontal length scale of motion. In this limit, hydrostatic balance holds in the vertical direction, and the horizontal velocity field  $\mathbf{v(\mathbf{r})}$  is depth independent. The dynamics then follows from mass and horizontal momentum conservation. Shallow water waves are solutions of this model linearized around a state of rest. In the case where the dynamics takes place on a flat surface, it can be written as the following Hermitian eigenvalue problem, quite similarly to a quantum mechanical problem
 \begin{align}
 \label{eq:SW}
\begin{pmatrix}
0 & -\ii f(y)  & k_x \\
\ii f(y) & 0  &  \ii \partial_y\\
k_x & \ii \partial_y & 0  
\end{pmatrix}
\begin{pmatrix}
\tilde{v}_x\\ \tilde{v}_y \\ \tilde{\eta}
\end{pmatrix}
= \tilde{\omega}(k_x)
\begin{pmatrix}
\tilde{v}_x\\ \tilde{v}_y \\ \tilde{\eta}
\end{pmatrix} .
\end{align}
The two horizontal velocity components $(\tilde{v}_x,\tilde{v}_y)$  are coupled to the free surface elevation  $\tilde{\eta}$. Time unit has been chosen so that the intrinsic celerity $c=\sqrt{gH}$ has been set to unity, with $g$ the standard gravity and $H$ the averaged layer thickness (see figure \ref{fig:SW} (a)). In Eq. \eqref{eq:SW}, $k_x$ is the eastward wave vector component while $y$ refers to the direction pointing to the North pole. The function $f(y)$ accounts for the variations with latitude of the Coriolis parameter.  
 
Eigenmodes of Eq. \eqref{eq:SW} are  tricky to derive analytically for an arbitrary profile $f(y)$, and becomes even more complicated when curvature effects are taken into account. A useful and standard simplification called the $\beta$-plane approximation consists in linearizing the Coriolis parameter around the Equator, i.e. $f(y) =\beta y$. This approximation was fruitfully used by Taroh Matsuno in 1966 to derive the frequency spectrum shown in figure \ref{fig:SW}(b) \cite{Matsuno1966}. This spectrum, besides being discrete, shows that the modes gather in two groups separated by a gap of amplitude $\sqrt{\beta c}$~: the low frequency Rossby waves and the high frequency Poincaré waves. Negative eigenvalues are also shown, for reasons that will become clear in the topological analysis below. These modes are simply related to the positive ones by a symmetry $\tilde{\omega}(k_x) \rightarrow -\tilde{\omega}(-k_x)$ which is analogous to the particle-hole symmetry in quantum mechanics. 

In addition to Rossby and Poincar\'e wave modes, Matsuno found two additional waves whose dispersions connect the two previous branches by continuously bridging the frequency gap. These two modes, called  equatorial Kelvin and Yanai waves, share the remarkable property to have a positive group velocity at any wave vector, so that they always propagate energy eastward, in contrast to Rossby and Poincaré waves. Moreover, there is a finite range of frequency in which these modes are the only one to be excited. This phenomenology is precisely that of the edge modes of a two-dimensional topological material where time-reversal symmetry is broken. In topological insulators, the topological properties emerge in the bulk, that is in a simplified edgeless problem where translation symmetry is restored. The situation is a bit different here, since the equatorial problem is already edgeless. However, a ''bulk Hamiltonian'' can also be assigned to the situation where  translation symmetry would be restored in the $y$ direction, by considering the Coriolis function varying with latitude as a (constant) parameter $f$, and then Fourier transform with respect to the $y$ coordinate. This leads to the distinct parametrized eigenvalue problem 
\begin{align}
\label{eq:SWf}
\begin{pmatrix}
0 & -\ii f  & k_x \\
\ii f & 0  &  k_y\\
k_x & k_y & 0  
\end{pmatrix}
\begin{pmatrix}
v_x\\ v_y \\ \eta
\end{pmatrix}
= \omega(k_x,k_y,f)
\begin{pmatrix}
v_x\\ v_y \\ \eta
\end{pmatrix}
\end{align}
where the eigenfrequencies  simply consist in three bands $ \omega_\pm = \pm \sqrt{k_x^2+k_y^2+f^2}$ and $\omega_0=0 $. Dispersion relations are shown for different values of $f$ in figure \ref{fig:SW}(c).  The modes  $\omega_\pm$ are separated to the flat one $\omega_0$ by a frequency gap of amplitude $|f|$. In other words, at fixed $f$, the system behaves as a two-dimensional insulator with broken time-reversal symmetry. In a condensed matter context, the standard next move would be to compute the Chern number of each of the bands, by integrating the Berry curvature -- obtained from the eigenstates of \eqref{eq:SWf} -- over the close manifold parameter space, in that case, the two-dimensional Brillouin zone span by $k_x$ and $k_y$, that is equivalent to a torus. However, for geophysical waves, and more generally for waves in continuous media, the absence of a Brillouin zone makes this procedure inappropriate. Instead of considering the waves for a single fixed value of $f$, one can look on the  continuous family of solutions in parameter space $(k_x,k_y,f)$. 

Three eigenfrequencies and their corresponding eigenvectors, that we shall note $\Psi_\pm$ and $\Psi_0$,  are assigned to each point of the parameter space. Once normalized, the eigenstates do not depend on their distance from the origin in parameter space, but only on the orientation. In other words, they live on a sphere that surrounds the origin. Note that the origin is a very particular point of the problem, since there the three bands touch. Bands touching points are known to be a source of Berry curvature which, once integrated over the sphere that surrounds it, yields a quantized Chern number $\mathcal{C}=\frac{1}{2\pi}\int_{S^2}\mathcal{F} \in \mathbb{Z}$. 
Surprisingly, this shallow water problem is formally analog to that of the quantum spin coupled to a classical magnetic field briefly introduced in section \ref{sec:geomtopo}. Indeed, the Hermitian matrix in \eqref{eq:SWf} playing the role of the quantum Hamiltonian takes the form $\mathbf{B}.\mathbf{\hat{S}}$ where $\mathbf{B}=(k_x,k_y,f)$ and $\mathbf{\hat{S}}$ is a spin-$1$ operator. The Chern numbers can then be inferred directly from the Berry phase when the solid angle is that of the entire sphere, i.e. $\mathcal{C}_m = \frac{1}{2\pi} (-m) 4\pi$  that gives $(2,0,-2)$ for the three bands \cite{delplace2017topological}.

Importantly, these Chern numbers are directly related to a spectral property of the equatorial frequency spectrum $\tilde{\omega}(k_x)$~: a band $n$ of this spectrum gains $-\mathcal{C}_n$ modes when increasing $k_x$, as emphasized in figure \ref{fig:SW}(b). It follows that modes have to transit from band to band; this is referred to as a \textit{spectral flow}. The set of Chern numbers found in the parameterized problem \eqref{eq:SWf} thus imposes that two modes of positive group velocity have to fill the gap, hence the topological origin of the Kelvin and the Yanai waves \cite{Faure}. This topological information does not depend on the shape of  the Coriolis function $f(y)$, which can be continuously modified without changing the spectral flow. 
The existence of these two unidirectional modes is thus guaranteed beyond the $\beta$ plane approximation.

 First observations of equatorial Kelvin and Yanai waves in Earth atmosphere and oceans came almost simultaneously to their theoretical predictions, in the late sixties and the seventies. Their discovery has been central to our modern understanding of tropical dynamics. Indeed, they are involved in the most important internal modes of climate variability in the equatorial area. For instance, equatorial Kelvin waves travelling across the Pacific ocean are precursors of El Ni\~no events, atmospheric equatorial Kelvin waves are often invoked in the explanation for superrotation phenomenon reported in other planets, including exoplanets, and the  Quasi-Biennial Oscillations of stratospheric winds are partly driven by atmospheric Yanai and equatorial  Kelvin waves in the middle atmosphere, see e.g. \cite{VallisBook}. 
 
\section{Conclusion and prospects}

Unidirectional equatorial waves can be predicted from topological considerations without solving the rotating shallow water model on the sphere, just as the deviation of the Foucault pendulum could be inferred from geometrical considerations only, without deriving the equations of motion. Besides, it highlights the important role of broken symmetries in a system at hand, and of interfaces induced by the symmetry breaking parameter.  
This topological approach is not restricted to the shallow water model, nor to systems subject to the Coriolis force. Similar arguments can be used to predict in a simple way the existence of remarkable waves in various complicated continuous systems such as astrophysical and geophysical flows. 
The strategy  consists in searching for band crossing points in dispersion relations,  identifying the three-dimensional parameter space in which these degeneracies occur, and  computing the topological charge describing eigenmodes twisting around these degeneracy points in parameter space. It is then possible to use these information to predict the emergence of trapped modes along an interface in physical space. Beyond the equatorial case, this method has already been used to predict the emergence of Lamb-like waves in compressible-stratified fluids \cite{perrot2019topological}, with possible applications in astroseismology,  or to peculiar plasma waves \cite{parker2020topological}, with possible experimental realizations.The discovery of  other topological waves in seismology, astrophysics, and geophysical fluid dynamics is at hand.

\bibliographystyle{unsrt}
\bibliography{bibliography}

\end{document}